\documentclass[final,5p,times,twocolumn]{elsarticle}

\usepackage{amssymb}
\usepackage{color}
\usepackage{xcolor}
\usepackage[active]{srcltx}
\usepackage{amsmath,amsfonts,amssymb,amsthm,amstext,amscd,eucal,srcltx}
\usepackage{epsfig,graphicx,bm}
\usepackage{epstopdf, epsf}
\usepackage{dcolumn}
\usepackage{comment}
\usepackage{hyperref}
\usepackage{float}
\usepackage{lipsum}
\usepackage[caption=false]{subfig}
\usepackage{xspace}
\usepackage[running]{lineno}
\usepackage{upgreek}
\usepackage{multirow}
\usepackage{verbatim}
\usepackage{ulem}

\newcommand{\be}{\begin{equation}}
\newcommand{\ee}{\end{equation}}

\newcommand{\bse}{\begin{subequations}}
\newcommand{\ese}{\end{subequations}}
\newcommand{\bea}{\begin{eqnarray}}
\newcommand{\eea}{\end{eqnarray}}
\newcommand{\ba}{\begin{array}}
\newcommand{\ea}{\end{array}}
\newcommand{\bc}{\begin{center}}
\newcommand{\ec}{\end{center}}

\newcommand{\kstar}        {\ensuremath{k^*}\xspace}
\newcommand{\thirteen}        {$\sqrt{s}~=~13$~Te\kern-.1emV\xspace}
\newcommand{\Ledn}         {Lednick\'y--Lyuboshits approach\xspace}
\newcommand{\Nphi}         {\ensuremath{\mathrm{N}}\mbox{--}\ensuremath{\phi}\xspace~}
\newcommand{\pphi}         {\ensuremath{\mathrm{p}}\mbox{--}\ensuremath{\phi}\xspace}

\journal{Physics Letters B}

\begin{document}

\begin{frontmatter}



\title{Indication of a p--\texorpdfstring{$\phi$}~ bound state from a correlation function analysis}

\author{Emma Chizzali$^{a,b}$}
\ead{emma.chizzali@tum.de}
\author{Yuki Kamiya$^{c,d}$}
\ead{kamiya@hiskp.uni-bonn.de}
\author{Raffaele Del Grande$^{b}$}
\author{Takumi Doi$^{d}$}
\author{Laura Fabbietti$^{b}$}
\author{Tetsuo Hatsuda$^{d}$}
\author{Yan Lyu$^{e,d}$ \\
\vspace{0.5 cm}
{\it$^a$ Max-Planck-Institut für Physik, D-80805 Munich, Germany, EU}\\
{\it$^b$ Physik Department E62, Technische Universität München, Garching, Germany, EU}\\
{\it$^c$ Helmholtz Institut für Strahlen- und Kernphysik and Bethe Center for Theoretical Physics, Universität Bonn, D-53115 Bonn, Germany}\\ 
{\it$^d$ RIKEN Interdisciplinary Theoretical and Mathematical Science Program (iTHEMS), Wako 351-0198, Japan}\\
{\it$^e$ State Key Laboratory of Nuclear Physics and Technology, School of Physics, Peking University, Beijing 100871, China}\\
\vspace{0.5 cm}
{IPM/P-2012/009, RIKEN-iTHEMS-Report-22, MPP-2022-290}  
}

\begin{abstract}
\noindent
The existence of a nucleon--$\phi$ (N--$\phi$) bound state has been subject of theoretical and experimental investigations for decades. In this letter, indication of a \pphi bound state is found, using for the first time two-particle correlation functions as alternative to invariant mass spectra. Newly available lattice calculations for the spin 3/2 \Nphi interaction by the HAL QCD collaboration are used to constrain the spin 1/2 counterpart from the fit of the experimental \pphi correlation function measured by ALICE. 
The corresponding scattering length and effective range are $f_0^{(1/2)}=\left(-1.54^{+0.53}_{-0.53}(\mathrm{stat.})^{+0.16}_{-0.09}(\mathrm{syst.})+i\cdot0.00^{+0.35}_{-0.00}(\mathrm{stat.})^{+0.16}_{-0.00}(\mathrm{syst.})\right)$~fm and $d_0^{(1/2)}=\left(0.39^{+0.09}_{-0.09}(\mathrm{stat.})^{+0.02}_{-0.03}(\mathrm{syst.})+i\cdot0.00^{+0.00}_{-0.04}(\mathrm{stat.})^{+0.00}_{-0.02}(\mathrm{syst.})\right)$~fm, respectively. The results imply the appearance of a \pphi bound state with an estimated binding energy in the range of $12.8-56.1$ MeV. 
\end{abstract}


\end{frontmatter}


One of the still debated and open questions in strangeness nuclear physics concerns the possible existence of $\phi$-mesic bound states with nucleons (N)~\cite{PhysRevC.63.022201,PhysRevC.95.055202,PhysRevC.75.058201,PhysRevC.73.025207,aliceBS,HIRENZAKI2010406,Yamagata-Sekihara:2010ell,PhysRevC.96.035201,Sofianos_2010,Belyaev_2008}. The main building block for the investigation of such a bound state is a precise knowledge of the \Nphi strong interaction. 
The latter can be theoretically investigated in the framework of the hidden gauge theory with unitary coupled-channel calculations within SU(3) symmetry~\cite{oset2010dynamically,Oset:2012ap,RAMOS2013287} or by employing a spin-flavor SU(6) extension of the SU(3) chiral perturbation theory~\cite{PhysRevD.84.056017}. Both types of calculations indicate a negligible elastic 
$\rm {N}\phi\rightarrow  \rm {N}\phi$ coupling, since the $\phi$ meson is mainly composed of $s\bar{s}$ and its interaction with nucleons is expected to be suppressed because of the Okubo-Zweig-Iizuka (OZI) rule~\cite{OZI_O,OZI_Z,OZI_I}.
If inelastic channels such as $\rm {N}\phi \rightarrow K^*\Lambda/\Sigma$ are taken into account in these calculations, the interaction can proceed via OZI allowed couplings. In \cite{oset2010dynamically,Oset:2012ap,RAMOS2013287,PhysRevD.84.056017} no \Nphi bound state is predicted, while in other chiral SU(3) models \cite{PhysRevC.73.025207,aliceBS} a bound state emerges. 
A different theoretical approach is to consider QCD van der Waals forces, where the \Nphi interaction is mediated by multi-gluon instead of quark exchange~\cite{QCDvdW}. Also in this case the interaction is found to be sufficiently attractive to support a \Nphi bound state~\cite{PhysRevC.63.022201}.

Experimentally, no evidence of a \Nphi bound state has been found yet, as standard invariant mass measurements of its decay products are challenging. Also the measurement of the \pphi scattering parameters are currently limited to spin averaged quantities.
The LEPS collaboration measured the differential cross sections of coherent $\phi$-meson photo-production off a deuteron target, with a photon beam energy of $E_{\gamma}=1.5-2.4$~GeV~\cite{LEPS}. The measurements were interpreted with the help of QCD sum rule calculations~\cite{PhysRevC.76.048202} and were found to be consistent with an average scattering length of $f_0\sim0.15$~fm. 
The CLAS collaboration measured the near-threshold total $\phi$ cross section in $\gamma\mathrm{p}\rightarrow \phi \mathrm{p}$
reactions with $E_{\gamma}=1.63-2.82$~GeV.
The experimental results were interpreted employing the vector meson dominance model (VMD) and a scattering length of $|f_0|=(0.063\pm0.010)$~fm was obtained~\cite{Strakovsky:2020uqs}. Recent measurements of the HADES collaboration showed, however, that by employing strict VMD, the measured dilepton yields can not be reproduced~\cite{VDM_hades}.

The most recent experimental result was obtained by the ALICE collaboration by measuring the proton-$\phi$ correlation function in momentum space in pp collisions at \thirteen~\cite{ALICEpphi}.
The data was modeled using the \Ledn~\cite{Lednicky:1981su} and a spin-averaged scattering length of $f_0=(0.85\pm0.34(\mathrm{stat.})\pm0.14(\mathrm{syst.})+i\cdot0.16\pm0.10(\mathrm{stat.})\pm0.09(\mathrm{syst.}))$~fm was extracted. 
Contrary to the previous measurements and theoretical expectations, this result indicates a larger scattering length $f_0$. It also hints to a negligible inelastic contribution to the \pphi interaction.

In addition, measurements of $\phi$ absorption off different nuclear targets in photon-~\cite{CLAS_absorp}, proton-~\cite{ANKE_absorp} and pion-induced reactions \cite{HADES:2018qkj} were carried out. All these measurements indicate a dominant inelastic contribution due to the absorption of $\phi$-mesons in nuclear matter.
In summary, the landscape of experimental results for the \Nphi interaction is not completely consistent and new inputs are necessary.

In this letter a re-analysis of the ALICE data is discussed, with the aim to disentangle the two spin components $(\frac{1}{2},\frac{3}{2})$ of the \pphi interaction. 
By considering the lattice calculation for the spin 3/2 channel 
by the HAL QCD collaboration~\cite{Lyu_Nphi_PRD2022}, the spin 1/2 interaction is constrained by fitting the \pphi correlation function measured by ALICE. The result supports the formation of a bound state. 

\begin{figure}
    \centering
    \includegraphics[width=8.6cm]{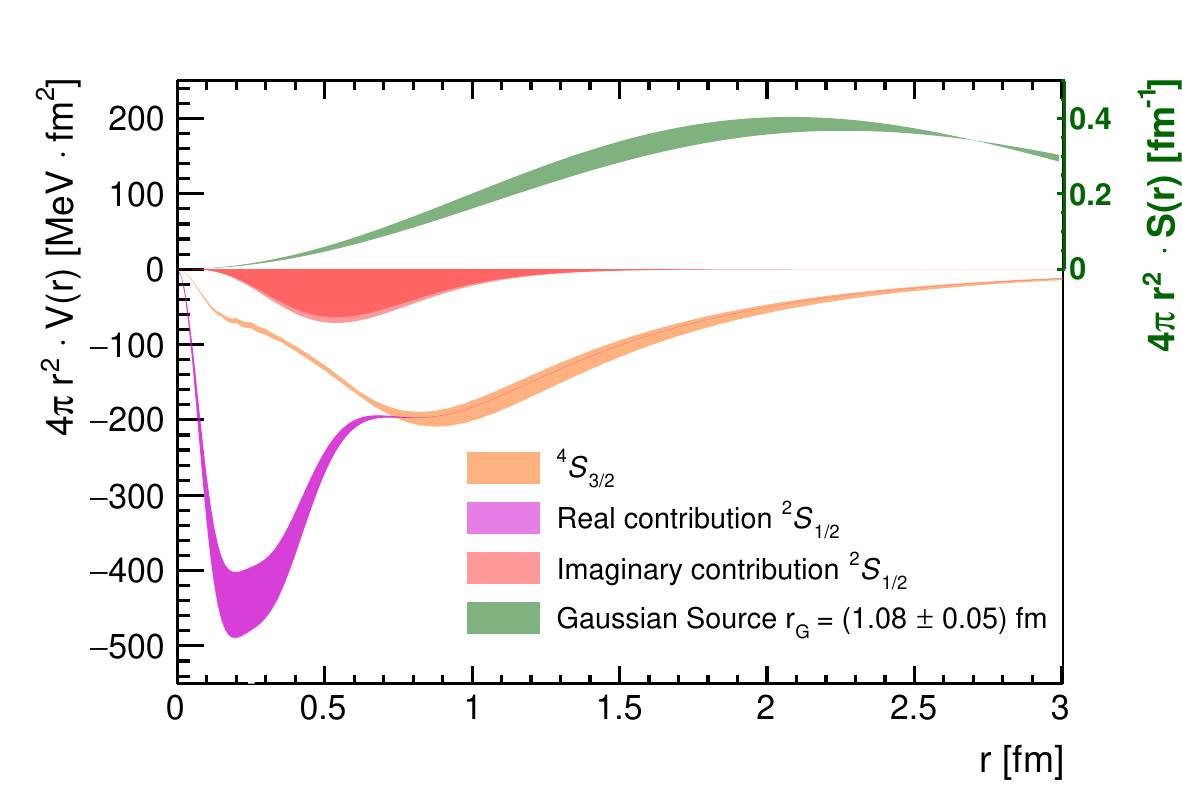}
    \caption{The parametrization of the \Nphi lattice potential for the $^4S_{3/2}$ channel at physical masses (orange band) is shown as a function of the hadron distance together with the real (violet band) and imaginary (red band) part of the complex phenomenological potential  for the $^2S_{1/2}$ channel obtained from the fit. Also the probability density distribution for the hadron distance
    $4\pi r^2 \cdot S(r)$ is depicted (green band). The potentials are multiplied by the Jacobian term $4\pi r^2$ to emphasize the range of sensitivity of the correlation function measurement to the potential. Dark bands correspond to the statistical error of the experimental correlation function and light band to the total uncertainty, including also the systematic uncertainties $\sigma_{\mathrm{tot}}=\sqrt{\sigma_{\mathrm{stat}}^2+\sigma_{\mathrm{syst}}^2}$.
    }
    \label{fig:pot}
\end{figure}

A recent lattice QCD calculation performed for the \Nphi interaction in the $S$-wave and spin 3/2  channel ($^4S_{3/2}$) is based on the (2+1)-flavor simulation with nearly physical quark masses $(m_\pi, m_K)\simeq(146, 525)~\text{MeV/}c^2$ and a large lattice volume
of $\simeq (8.1~\text{fm})^3$~\cite{Lyu_Nphi_PRD2022}. 
The obtained interaction potential from the lattice measurement of \Nphi space-time correlation~\cite{Ishii2007,Ishii2012} can be converted in a scattering length of 
$f_0^{(3/2)}= -a_0^{(3/2)} =1.43 \pm 0.23\text{ (stat.)}^{+0.06}_{-0.36}\text{ (syst.)}$ fm and an effective range of 
$d^{(3/2)}_0 = r_{\rm eff}^{(3/2)} =2.36\pm0.10\text{ (stat.)}^{+0.02}_{-0.48}\text{ (syst.)}$ fm.
The N$\phi (^4S_{3/2})$ potential is found to be a combination of an attractive core at short distances and a two-pion-exchange (TPE) tail at long distances.
In particular, the latter is consistent with the characteristic form of the TPE obtained by the interaction of a color-dipole and the nucleon~\cite{Castella2018}.
By examining the $m_\pi$-dependence of the TPE tail, it is suggested that the attractive potential could be slightly weaker at the physical point, $m_\pi = 138$~MeV/$c^2$.

The coupling of N$\phi(^4S_{3/2})$ with $ \Lambda \mathrm{K}(^2D_{3/2})$ and $\Sigma \mathrm{K}(^2D_{3/2})$ is kinematically suppressed at low energies due to their $D$-wave nature and hence the effect from these coupled channels is invisible in the lattice calculation. The  N$\phi(^2S_{1/2})$ state is different since it can couple via $S$-wave to the $\Lambda \mathrm{K}(^2S_{1/2})$ and $\Sigma \mathrm{K}(^2S_{1/2})$ states~\cite{Lyu_Nphi_PRD2022}.
Such decay processes can be described through the exchange of kaons, and this introduces an imaginary part for the potential in the N$\phi(^2S_{1/2})$ channel.
On the other hand, the long-range TPE potential found in the N$\phi(^4S_{3/2})$ channel should also characterize the N$\phi(^2S_{1/2})$ channel, since the exchange of two pions in a scalar-isoscalar state (similar to the $\sigma$-meson exchange) does not depend on the total spin of the N$\phi$ system.
Therefore, the diagonal N$\phi(^2S_{1/2})$ potential is expected to consist of a long-range attractive TPE potential as for the N$\phi(^4S_{3/2})$ channel combined with a phenomenological term at short distances, and an imaginary potential characterized by a 2nd-order kaon exchange.
 These considerations motivate the following  potential form for the N$\phi(^2S_{1/2})$ channel
\begin{align}
 V_\mathrm{N\phi}(r) &= \beta \left( \sum_{i=1,2}a_i e^{-(r/b_i)^2} \right) 
 + a_3 m_\pi^4 f(r;b_3)  \frac{e^{-2m_\pi r}}{r^{2}} \nonumber \\
 &+ i \gamma  f(r;b_3)  \frac{e^{-2m_\mathrm{K} r}}{m_\mathrm{K} r^2},
 \label{eq:s12}
\end{align}
where $\beta$ and $\gamma$ are free parameters to be constrained by the fit to experimental data; $a_{1,2,3}$ and $b_{1,2,3}$ are obtained by the fit to the lattice QCD data in the spin 3/2 channel at $t/a=12$\footnote{This value was chosen over the default $t/a=14$ in~\cite{Lyu_Nphi_PRD2022} to be conservative, as it corresponds to the least attractive potential in the spin 3/2 channel, hence, also to the weakest potential in the spin 1/2.} provided in~\cite{Lyu_Nphi_PRD2022} 
and are common in both $^4S_{3/2}$ and $^2S_{1/2}$ channels, whose values are listed in Tab.~\ref{tab:ta12parval}; 
$f(r;b_3) = \left( 1-e^{-(r/b_3)^{2}} \right)^2$ is the Argonne-type form factor, also discussed in~\cite{Lyu_Nphi_PRD2022}. 
The $\beta$ parameter being positive for short-range attraction and negative for short-range repulsion can assume any real number,
while $\gamma$ is restricted only to negative values that correspond to absorption processes.
\begin{table}[]
    \centering
    \begin{tabular}{c|c}
        Parameter & Value \\
        \hline\hline
        $a_1$ [MeV]& -392(10)\\
        $b_1$ [fm]& 0.128(3)\\
        $a_2$ [MeV]& -145(9)\\
        $b_2$ [fm]& 0.284(7)\\
        $a_3m_{\pi}^4$ [MeV $\cdot$ fm$^2$]& -83(1)\\
        $b_3$ [fm]& 0.582(6)\\
        \hline\hline
    \end{tabular}
    \caption{Parameter values obtained by a fit to the lattice potential for the spin 3/2 channel at $t/a=12$~\cite{Lyu_Nphi_PRD2022}. }
    \label{tab:ta12parval}
\end{table}

For $\beta=1$ and $\gamma=0$, the potential in Eq.~(\ref{eq:s12}) is equivalent to the lattice QCD potential obtained in~\cite{Lyu_Nphi_PRD2022} for the spin 3/2 channel.
In order to properly model the experimental correlation function, physical masses are used for both $m_\pi$ and $m_\mathrm{K}$. The orange band in Fig.~\ref{fig:pot} shows the spin 3/2 potential function, using $(m_\pi, m_\mathrm{K}) = (138, 496)$~MeV/$c^2$, while all other parameters in Eq.~(\ref{eq:s12}) are 
equivalent to the parameters extracted from the original lattice data at nearly physical quark masses.

The $\beta$ and $\gamma$ parameters for the spin 1/2 channel are determined via a fit to the experimental \pphi correlation function, measured by the ALICE collaboration in pp collisions at \thirteen~\cite{ALICEpphi}. This observable has been used successfully to study various two-body interactions~\cite{FemtoRun1, FemtoSource, FemtoLambdaLambda, FemtoKaon, FemtopXi, FemtopSigma, FemtoNature, FemtoRev, LXIGeorgios, pLambdaALICE, BantiB,ALICE:2023wjz} and was recently extended to the three-body sector~\cite{DelGrande:2021mju,ALICE:2022vzr,alicecollaboration2023study}.

Following~\cite{Lisa:2005dd}, the correlation function is defined as
\begin{equation}
C(k^*)= \int \, d^3 r^{*} S(r^*) |\Psi(\vec{r}^{ \,*},\vec{k}^{*})|^2,
\label{eq:CFsourcewf}
\end{equation}where $\vec{k}^{\,*}=\frac{1}{2}\cdot ( \vec{p_1}^*-\vec{p_2}^* )$ is the reduced relative momentum of the pair of interest in its rest frame, denoted by ($^*$), $\vec{p_i}^*$ the momentum of the particle $i$, and $r^*$ the relative distance between the production points of the two particles. Moreover, $\Psi(\vec{r}^{ \,*},\vec{k}^{*})$ represents the relative wave function of the particle pair, which incorporates the final-state interaction, and $S(r^*)$ the source function, which describes the probability of emitting a pair at relative distance $r^*$~\cite{FemtoRev}. 
The wave function $\Psi(\vec{r}^{ \,*},\vec{k}^{*})$ must satisfy the outgoing boundary condition where the flux of the outgoing wave is normalized.
The source function $S(r^*)$ is modeled by a Gaussian distribution, using a source radius of $r_{G}=(1.08\pm0.05)$~fm~\cite{FemtoSource,ALICEpphi}. This value is anchored to the measured transverse mass of the \pphi pair and also accounts for secondary protons stemming from the strong decay of short live resonances that lead to an increase of the source radius~\cite{FemtoSource}. The resulting probability density distribution $4\pi r^{*2}\cdot S(r^*)$ is shown by the green band in Fig.~\ref{fig:pot}.
\begin{figure}
    \centering
    \includegraphics[width=8.6cm]{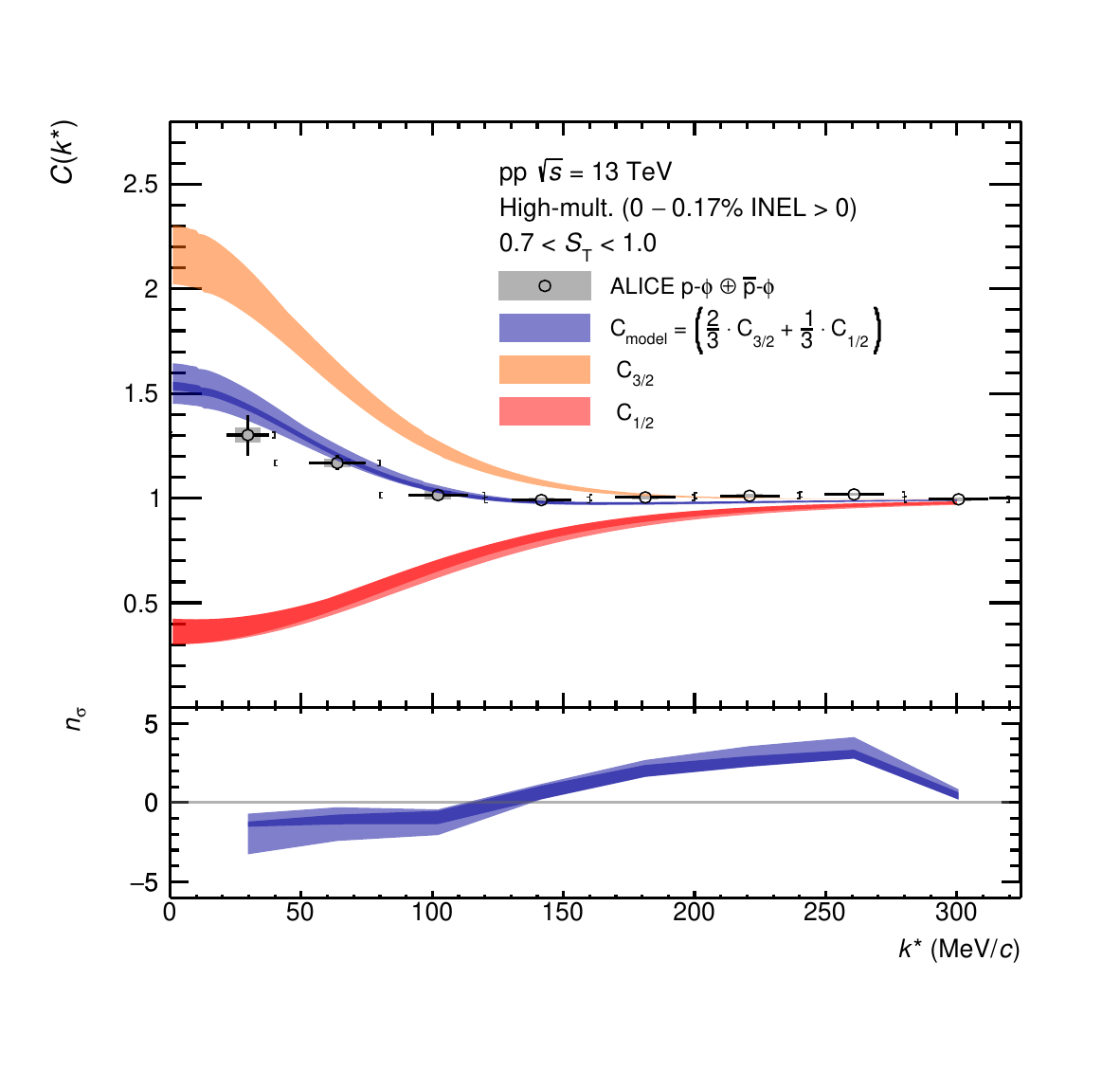}
    \caption{The experimental \pphi correlation function measured by the ALICE collaboration~\cite{ALICEpphi} is depicted in the upper panel with systematic (gray shaded squares) and statistical uncertainties (lines) together with the spin averaged model correlation function (blue band) and the unweighted $^{4}S_{3/2}$ (orange band) and $^{2}S_{1/2}$ contributions (red band). The dark shaded bands arise to the uncertainty from the statistical error of the ALICE data, while the light shaded ones correspond to the total error, which includes the systematic uncertainty via $\sigma_{\mathrm{tot}}=\sqrt{\sigma_{\mathrm{stat}}^2+\sigma_{\mathrm{syst}}^2}$. The lower panel shows the number of standard deviations $n_{\sigma}$ between $C_{\mathrm{model}}$ and data.
    }
    \label{fig:cf}
\end{figure}

\begin{figure}
    \centering
    \includegraphics[width=8.6cm]{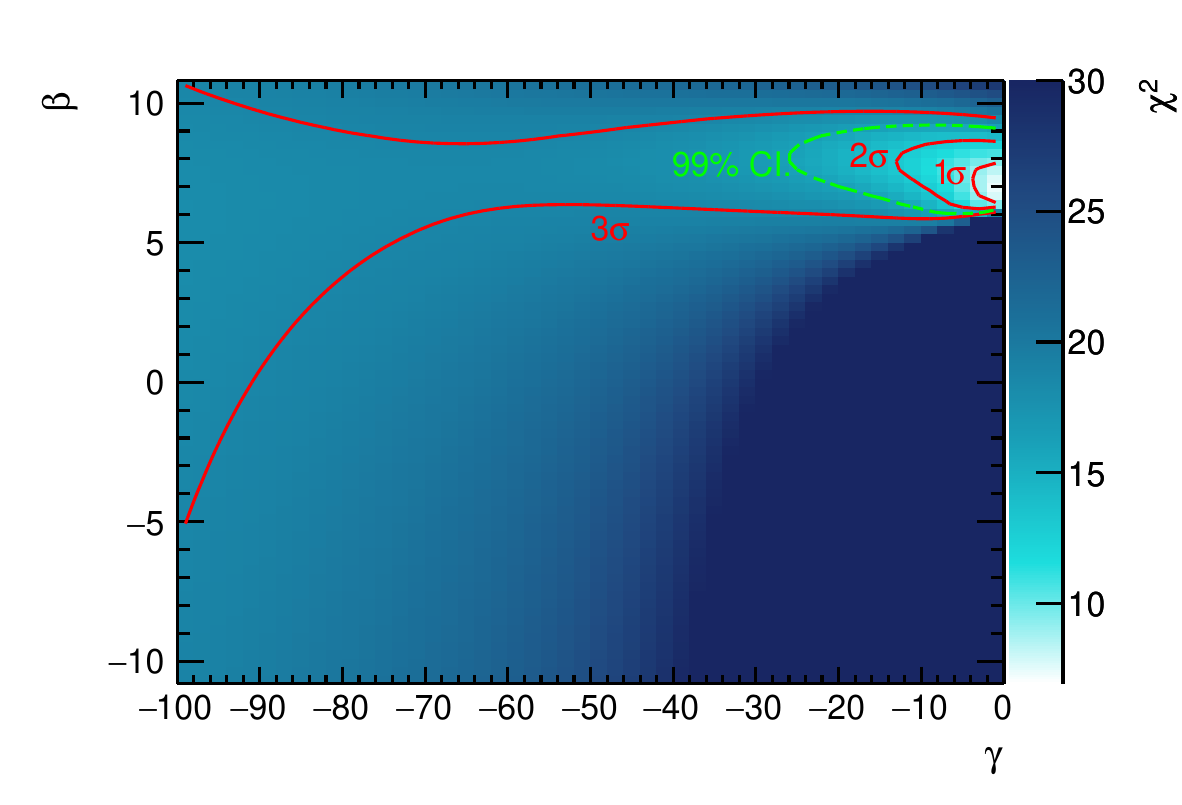}
    \caption{Distribution of $\chi^2$ in the ($\beta,\gamma$) plane, where $\beta$ ($\gamma$) parameter defined in Eq.~(\ref{eq:s12}) controls the strength of the real (imaginary) part of the $^{2}S_{1/2}$ potential. The $\chi^2$ is evaluated within $k^*\in[0,200]$~MeV/$c$, considering statistical uncertainties of the data. The red lines correspond to the 1-, 2- and 3-$\sigma$ contour with respect to the minimum $\chi^2_{\mathrm{min}}=6.85$ ($n_{\sigma}=1.77$) at $\beta=6.9$ and $\gamma=0.0$, the green dashed line shows the 99\% Cl..}
    \label{fig:phasespace}
\end{figure}

At small relative momentum $k^*<200$~MeV/$c$, typically referred to as femtoscopic region, the genuine correlation function is sensitive to the final state interaction, as the two particles are close enough in momentum space to interact with each other. $C(k^*)>1$ in the femtoscopic region usually corresponds to an attractive interaction, while $C(k^*)<1$ is caused by either a repulsive interaction or an attractive interaction which is strong enough to support the formation of a bound state. In case of no interaction, the genuine correlation function is flat and $C(k^*)=1$, which can also be observed at high relative momenta, $k^*\rightarrow \infty$, as the two particles separate fast enough to avoid any final state interaction. Details can be found in~\cite{FemtoRev}. The histogram in Fig.~\ref{fig:cf} shows the experimental \pphi correlation function measured by ALICE. The orange band in the same figure shows the calculated correlation function  obtained considering only the $^4S_{3/2}$ channel.
The latter is evaluated with Eq.~(\ref{eq:CFsourcewf}) and employing the CATS framework~\cite{CATS} 
to obtain the relative \pphi wave function starting from the parametrization of the published lattice potential~\cite{Lyu_Nphi_PRD2022}. One can see that the $^4S_{3/2}$ channel alone overestimates the experimental data. Since the \pphi system is characterized by the two spin states $(\frac{1}{2},\frac{3}{2})$, the total correlation function reads as
\begin{equation}
C_{\rm model}^{(\beta,\gamma)}(k^*)=\frac{2}{3}C_{3/2}(k^*)+\frac{1}{3}C_{1/2}^{(\beta,\gamma)}(k^*) \ ,
\end{equation} 
where the dependence from the free $(\beta,\gamma)$ parameters is explicitly indicated. The weight of each spin contribution is provided by the corresponding spin multiplicity. 
The spin 3/2 contribution is fixed by the HAL QCD prediction, while in order to constrain the spin 1/2 contribution, a minimum $\chi^2$ study is performed by varying the $\beta$ and $\gamma$ parameters of the complex potential given by Eq.~(\ref{eq:s12}).
To obtain the wave function that fulfills the outgoing boundary condition of the complex optical potential, the complex conjugate of $V_\mathrm{N\phi}(r)$ is used when solving the Schr\"odinger equation (for more details see Appendix A in~\cite{Kamiya:2022thy}).
For each $\beta$ and $\gamma$ combination the $C_{1/2}^{(\beta,\gamma)}(k^*)$ correlation function is computed and $C_{\rm model}^{(\beta,\gamma)}(k^*)$ is compared to the data.
The $\chi^2$ is defined as 
\begin{equation}
       \chi^2(\beta,\gamma) = \sum_{j=1}^{N} 
\left(
\frac{C_{\rm data}(k_j^*) - C_{\rm model}^{(\beta,\gamma)}(k_j^*)}{\sigma_{\rm data}(k_j^*)} \right)^2 ,
\end{equation}
where $N=5$ is the number of data points in the femtoscopic region $k^* < 200$ MeV/$c$, and $ \sigma_{\rm data}(k_j^*) $ is the uncertainty of the $j$-th ALICE data point. The $\chi^2$ distribution obtained considering only the statistical uncertainties of the measured correlation function is shown in Fig.~\ref{fig:phasespace}. The step-size for $\beta$ is 0.1 and for $\gamma$ 0.2. The red lines correspond to the 1-,2- and 3-$\sigma$ contours for the ($\beta,\gamma$) parameters with respect to the minimum $\chi^2$ and the green dashed line shows the 99\% CI..

The systematic uncertainty on the $^2S_{1/2}$ potential parameters is obtained from varying the source radius $r_{G}$ as well as the lattice potential within its uncertainties (statistical errors as well as systematic errors from Euclidean time dependence $t/a=12,13,14$ in the $^4S_{3/2}$ channel~\cite{Lyu_Nphi_PRD2022}), 
the upper limit of the $\chi^2$ evaluation interval by $\pm 50$ MeV/$c$, which changes the number of degrees of freedom by $\pm 1$.
For each variation the best-fitting $\beta$ and $\gamma$ values are extracted and the systematic errors are obtained from the width of the resulting parameter distributions.

The parameters of the $^2S_{1/2}$ potential are found to be
$\beta=6.9^{+0.9}_{-0.5}(\mathrm{stat.})^{+0.2}_{-0.1}(\mathrm{syst.})$ and $\gamma=0.0^{+0.0}_{-3.6}(\mathrm{stat.})^{+0.0}_{-1.8}(\mathrm{syst.})$.
The corresponding real (violet band) and imaginary (red band) part of the $^2S_{1/2}$ potential  multiplied by the Jacobian factor are shown in Fig.~\ref{fig:pot}. One can see that even though the best fit corresponds to a vanishing $\gamma$ value, a sizable ${\rm Im} V(r^*)_{1/2}$ can not be excluded. This effect is particularly evident at distances $r^*\sim1$~fm where a considerable amount of \pphi pairs are emitted. 
By comparing the real part of the $^2S_{1/2}$ with the $^4S_{3/2}$ potential, one can see that they overlap for $r^*>1$~fm, due to the shared TPE contribution, while for short distances the large $\beta$ value obtained from the fit leads to a more attractive potential in the $^2S_{1/2}$ case. Even though the probability density function peaks around $r^*\sim2$~fm, the influence of the potential at small $r^*$ is still notable and the region of $\beta<0$ is excluded by more than $3\sigma$ in the largest part of the phase-space down to a region of $\gamma\approx-90$ and within 99\% CI. in the full tested region.
The scattering length $f_0$ and effective range $d_0$  of the $^2S_{1/2}$ channel, extracted from the potential phase-shift, is found to be
\begin{equation}
\begin{aligned}
\label{eq:sc_parameters}
{\rm Re}\ f^{(1/2)}_0 & =-1.54^{+0.53}_{-0.53}(\mathrm{stat.})^{+0.16}_{-0.09}(\mathrm{syst.}) \ {\rm fm} ,  \\
{\rm Im}\ f^{(1/2)}_0 & = \ \   0.00^{+0.35}_{-0.00}(\mathrm{stat.})^{+0.16}_{-0.00}(\mathrm{syst.})\   {\rm fm} ,  \\
{\rm Re}\ d^{(1/2)}_0 & = + 0.39^{+0.09}_{-0.09}(\mathrm{stat.})^{+0.02}_{-0.03}(\mathrm{syst.}) \  {\rm fm} ,  \\
{\rm Im}\ d^{(1/2)}_0 & = \ \  0.00^{+0.00}_{-0.04}(\mathrm{stat.})^{+0.00}_{-0.02}(\mathrm{syst.}) \  {\rm fm} .
\end{aligned}
\end{equation}
The resulting model correlation function as well as the correlation functions of the individual spin contributions are shown in Fig.~\ref{fig:cf}. 
A good agreement with data is obtained. Considering the total uncertainty on both data and model leads to $\chi^2=3.44$ and $n_{\sigma}=0.98$ within $k^*\in[0,200]$~MeV/$c$.

Since the correlation function corresponding to the $^4S_{3/2}$ channel alone overshoots the data, the $^2S_{1/2}$ channel must be characterised either by a repulsive short-range real part of the potential with attractive pocket due to the TPE ($\beta<0$) or a fully attractive one ($\beta>0$). As can be seen in Fig.~\ref{fig:phasespace}, the best fit to the data is obtained for $\beta>0$, which corresponds to an attractive real part of the potential in the full $r^*$ range, as depicted in Fig.~\ref{fig:pot} while negative $\beta$ values are excluded within 99\% CI.. Therefore, this re-analysis of the experimental data indicates the formation of a \pphi bound state.
Other forms of the spin 1/2 and spin 3/2 potentials have been investigated by modifying the form factors according to the Nijmegen-type form factor $f_\mathrm{erfc}$ defined in Eq. (9) of~\cite{Lyu_Nphi_PRD2022}. The obtained spin 1/2 scattering parameters are found to be in agreement with those extracted from the standard fit and the $\beta$ parameter is always significantly larger than zero. Additionally, the sensitivity to the strongly attractive short-range part of the spin 1/2 potential was studied by setting $a_1=0$ in Eq.~(\ref{eq:s12}), which leads to a potential weaker by a factor of $\sim$ 4 for $r^*<0.3$~fm. The resulting scattering parameters are consistent with Eq.~(\ref{eq:sc_parameters}), and the best fit to the data is found for $\beta=10.4$ and $\gamma=0$. Finally, a simple double-Gaussian form of the real part of the spin 1/2 potential was tested by setting $a_{3}=0$ in Eq.~(\ref{eq:s12}). Hence, ${\rm Re} \ V(r^*)_{1/2}$ is attractive or repulsive in the full $r^*$-range, without the TPE tail found in the spin 3/2 channel~\cite{Lyu_Nphi_PRD2022}. The scattering parameters are consistent with the results obtained from the standard fit with TPE tail, and the best-fitting potential is characterised by $\beta=7.6$ and $\gamma=0$. This demonstrates the stability of the results and confirms the possible existence of a \pphi bound state below threshold.

Solving the Schr\"{o}dinger equation with Eq.(\ref{eq:s12}) leads to an eigenenergy of $E=-23.8^{+10.7}_{-32.2}(\mathrm{stat.})^{+2.7}_{-2.8}(\mathrm{syst.})-i\cdot0.0^{+0.0}_{-16.4}(\mathrm{stat.})^{+0.0}_{-6.6}(\mathrm{syst.})$~MeV. The real binding energy $E_{\rm B}$ corresponds to $-\mathrm{Re}\ E$, which leads to $E_{\rm B}=12.8$-$56.1$~MeV. An alternative approach is to employ the approximate formula~\cite{Naidon_2017} with the real values of $f_0$ and $d_0$ in Eq.~(\ref{eq:sc_parameters}) and the reduced mass $\mu$
\begin{equation}
  E_{\rm B}\simeq \frac{1}{2\mu d_0^2}\left(1-\sqrt{1+2\frac{d_0}{f_0}}\right)^2,
\end{equation}which results in $E_{\rm B}\simeq 10.7$-$120.5$ MeV, considering the total uncertainty on the scattering parameters.

\begin{table}[]
    \centering
    \begin{tabular}{c|c}
        $J^P$ & $E_{\rm{B}}$ [MeV]\\
        \hline\hline
        \multirow{2}{*}{$\frac{1}{2}^-$,$\frac{3}{2}^-$} & 1.8~\cite{PhysRevC.63.022201}\\
        & 9.0~\cite{aliceBS}\\
        & 9.3($\phi$n), 9.23($\phi$p)~\cite{Belyaev_2008}\\
        & 9.47~\cite{Sofianos_2010}\\
        \hline
        $\frac{1}{2}^-$ & 1.0-3.0~\cite{PhysRevC.73.025207}\\
        \hline
        \multirow{2}{*}{$\frac{3}{2}^-$} & 6.0-9.0~\cite{PhysRevC.73.025207}\\
        & 1.6-10.1~\cite{PhysRevC.95.055202}\\
        \hline\hline
    \end{tabular}
    \caption{Predicted binding energies of a proton--$\phi$ bound state with spin-parity $J^P$. The binding energy obtained from spin-independent potentials applies for both $\frac{1}{2}^-$ and $\frac{3}{2}^-$.
    }
    \label{tab:EBpredicitons}
\end{table}
These values are comparable or even larger than previous model calculations, which can be found in Tab.~\ref{tab:EBpredicitons}. A spin-independent QCD van der Waals attractive potential, modeled by a Yukawa-type of potential leads to a binding energy of 1.8 MeV~\cite{PhysRevC.63.022201}. Studies employing extended SU(3) chiral quark models lead to values of 1-3 MeV for spin-1/2, while much larger values of 6-9 MeV are obtained for spin-3/2 by taking into account the coupling to $\Lambda\mathrm{K}^*$~\cite{PhysRevC.73.025207}. 
Employing the resonating-group method within the quark delocalization color screening model a \Nphi bound state is found with main component $J^P=\frac{3}{2}^{-}$ and mass between 1948.9 and 1957.4~MeV/$c^2$, 
corresponding to a binding energy in the range of $[1.6,10.1]$~MeV~\cite{PhysRevC.95.055202}. 
Using unitary coupled-channel approximation anchored to the scattering length measured by ALICE~\cite{ALICEpphi}, a pole at 1949~MeV/$c^2$ is found~\cite{aliceBS}, which should correspond to a \Nphi bound state with $E_{B}=9$~MeV.
Using a phenomenological model following~\cite{PhysRevC.63.022201}, sizable binding energies of 9.3 for $\phi$--n and 9.23~MeV for $\phi$--p are obtained in~\cite{Belyaev_2008} by employing a variational method, while in~\cite{Sofianos_2010} an value of 9.47~MeV is found for \Nphi in a similar approach. Extending this approaches to A-body systems leads to much larger values of $E_{(\phi\mathrm{n)p}}=10.03$~MeV and $E_{(\phi\mathrm{n)p}}=17.45$~MeV in~\cite{Belyaev_2008} or up to $E_{\phi\mathrm{NN}}=39.84$~MeV and $E_{\phi\phi\mathrm{NN}}=124.59$~MeV in~\cite{Sofianos_2010}.
These results motivate a direct search for the \pphi bound state, which can decay for example into $\Lambda$K$^+$ or $\Sigma$K$^+$, both fall-apart decays in the S-wave. Facilities like J-PARC could measure the $\Lambda$K$^+$ channel for example in pA collisions. Also JLab is able to access it and probe a possible \pphi bound state. The ALICE Collaboration recently published the $\Lambda$K$^+\oplus\overline{\Lambda}$K$^-$ correlation function~\cite{ALICE:2023wjz}, where the \pphi bound state would appear as resonance structure in the region $\kstar\in[479, 517]~$MeV$/c$, taking into account the binding energy obtained in this work. However, no significant structure is observed in this region and other decay channels might be favoured. Also, no cusp at the opening of the  \pphi channel is found at $\kstar\sim529$~MeV$/c$.

In summary, in this letter a re-analysis of the experimental \pphi correlation function measured by ALICE~\cite{ALICEpphi} was performed, constraining the spin 3/2 channel with the recently published lattice QCD data~\cite{Lyu_Nphi_PRD2022}. This approach allowed to extract the spin 1/2 channel for the first time using
a phenomenological complex potential, whose shape is motivated by the lattice QCD calculation. 
The real part of the potential is found to be attractive and supports the existence of a \pphi bound state predicted in previous phenomenological calculations~\cite{PhysRevC.63.022201,PhysRevC.73.025207,PhysRevC.95.055202,aliceBS}. The imaginary part of the potential is vanishing, however, within uncertainties, it does not exclude the possibility of inelastic contributions expected by theory~\cite{oset2010dynamically,Oset:2012ap,RAMOS2013287,PhysRevD.84.056017} and experimental measurement off different nuclear targets~\cite{CLAS_absorp,ANKE_absorp,HADES:2018qkj}.

The obtained scattering length is $f_0^{(1/2)}=\left(-1.54^{+0.53}_{-0.53}(\mathrm{stat.})^{+0.16}_{-0.09}(\mathrm{syst.})+i\cdot0.00^{+0.35}_{-0.00}(\mathrm{stat.})^{+0.16}_{-0.00}(\mathrm{syst.})\right)$~fm and the effective range is $d_0^{(1/2)}=\left(0.39^{+0.09}_{-0.09}(\mathrm{stat.})^{+0.02}_{-0.03}(\mathrm{syst.})+i\cdot0.00^{+0.00}_{-0.04}(\mathrm{stat.})^{+0.00}_{-0.02}(\mathrm{syst.})\right)$~fm.
The binding energy of the \pphi bound state in the spin 1/2 channel is found to be in the range between $12.8$ and $56.1$ MeV.
Coupled channel lattice QCD calculations are called for to pin down more precisely the binding energy and the width of the \pphi system in the spin 1/2 channel.

The analysis presented in this letter demonstrates that correlation functions can be used to study the possible existence of bound states, as an alternative to the invariant mass technique. 
This stimulates future femtoscopic analyses aiming to search for more exotic bound states such as those involving charm hadrons.

\section*{Acknowledgements}
The authors thank Akira Ohnishi and Philipp Gubler for useful discussions.  This work was supported  by the ORIGINS cluster DFG under Germany’s Excellence Strategy - EXC2094 - 390783311 and the DFG  through Grant SFB 1258 “Neutrinos and Dark Matter in Astro- and Particle Physics”.
Y.K. was supported by the National Natural Science Foundation of China (NSFC) and the Deutsche Forschungsgemeinschaft (DFG) through the funds provided to the Sino-German Collaborative Research Center TRR110 ``Symmetries and the Emergence of Structure in QCD'' (NSFC Grant No.~12070131001, DFG Project-ID~196253076).
T.H. was supported by the JSPS grant JP18H05236. 
Y.L. was supported by the National Key R\&D Program of China (2017YFE0116700).
T.D. was supported by the JSPS grant JP18H05236, JP18H05407, JP19K03879, JP23H05439.
This work is supported in part by “Program for Promoting Researches on the Supercomputer Fugaku" (Simulation for basic science: from fundamental laws of particles to creation of nuclei), implemented by the Ministry of Education Culture Sports Science and Technology of Japan, and Joint Institute for Computational Fundamental Science (JICFuS). 

\bibliographystyle{elsarticle-num} 
\bibliography{bibliography}

\end{document}